\documentclass[aps,amsmath,amssymb,twocolumn,showpacs]{revtex4}
\usepackage{graphicx,color}
\graphicspath{{./}{./figures/}}
\usepackage{dcolumn}
\usepackage{bm}
\usepackage{subfigure}
\usepackage[english]{babel}

\begin{document}

\title{The balance of growth and risk in population dynamics}
\author{Thomas Gueudr\'e, David Martin} \affiliation{DISAT, Politecnico Corso Duca degli Abruzzi, I-10129 Torino - Italy} 
\date{\today\ -- \jobname}

\begin{abstract}
Essential to each other, growth and exploration are jointly observed in populations, be it alive such as animals and cells or inanimate such as goods and money. But their ability to move, crucial to cope with uncertainty and optimize returns, is tempered by the space/time properties of the environment. We investigate how the environment shape optimal growth and population distribution in such conditions. We uncover a trade-off between risks and returns by revisiting a common growth model over general graphs. Our results reveal a rich and nuanced picture: fruitful strategies commonly lead to risky positions, but this tension may nonetheless be alleviated by the geometry of the explored space. The applicability of our conclusions is subsequently illustrated over an empirical study of financial data.     
\end{abstract} 
\maketitle

Mingled search and collect of resources are central for sustainability. The exponential (or logistic) growth, most routinely observed in animals and cell populations \cite{royama2012analytical}, or in non biological contexts such as investment portfolios \cite{bouchaud2003theory}, has its instantaneous rate (or return) slaved to the available goods. As those fluctuate or get depleted, growth is often coupled with explorative dynamics that allow populations to forage new spots: humans create new cities, cell colonies spread or mutate, bankers reinvest profits and so on. To cope with uncertainty, they develop various strategies that strike a balance between costs and profits  \cite{PhysRevLett.119.140603,PhysRevLett.113.238101,PhysRevLett.108.088103,benichou2017optimally}. In such situations, many have been found to follow optimal policies, given their limitations in memory, mobility and sensing cues \cite{diaz2016global,ibarra2002escherichia,caraglio2016export}. 

At the heart of this optimization problem lie two timescales, set by the variability of the environment and the exploration speed. If a fertile spot remains so only for a finite duration, neither staying on it for long (and seeing it depleted), nor leaving it very rapidly (and failing to harvest it properly) provide optimal strategies for growth. Thus populations should tune their explorative behaviours according to this environmental timescale. However, populations growing exponentially are very sensitive to fluctuations and may fall in a localized phase, where they end up very concentrated on few abundant patches. This phenomenon is observed for example in interacting particles systems \cite{gärtner2010}, tumoral growth \cite{korolev2014turning} or portfolio rebalancing \cite{bouchaud2003theory}. Such condensation is commonly seen as undesirable, as putting all eggs in one basket reduces the system's resilience to shocks. The onset of localization is therefore frequently monitored by diversity indices, in ecology \cite{simpson1949measurement} or in economy \cite{federal2014department}.  

The joint concern of optimization and localization, quite familiar to finance and ecology, has also recently surfaced in the medical field, intervening in the optimal frequency of treatment administration, against rapidly mutating tumors \cite{Gatenby4894}. But in spite of some shared ingredients, the underlying space's geometry is specific to each case: animal populations mostly move on a plane, tumor cells jump over the genotype space, while profits in portfolio usually get redistributed over all assets. In front of this array of problems, natural questions arise. How do the fluctuations and structure of the environment affect growth ? What are the networks that best balance risks and returns ? Is there an optimal strategy ? If so, does it lead to a localized or well-spread population ?  

The aim of this Letter is to clarify how the environment, through stochasticity and geometry, shape those optimal strategies. We first revisit a well-known model, whose analysis we extend to colored noise and over general graphs. We extract several scalings and asymptotics that quantify the way diffusive populations grow over networks. The conclusions we derive from this model hold generally and allow for a principled approach to the risks-and-returns puzzle, as we illustrate by an empirical study over financial datasets. 

Our study model is the \textit{Parabolic Anderson Model} (PAM), a workhorse of stochastic exponential growth:
\begin{align}
\frac{\partial \phi(x_i,t)}{\partial t}  &= \lambda \Delta \phi(x_i,t) + \phi(x_i,t) \nu(x_i,t)  \label{PAM}
\end{align}
with the initial condition $\phi(x_i,0) = 1$ for all nodes $x_i$ in some graph $\mathcal{S}$. The exploration part here is diffusive, and written as the (normalized) discrete Laplacian operator $\Delta \phi(x_i,t) = |\mathcal{N}_{x_i}|^{-1} \sum_{j \in \mathcal{N}_{x_i}} (\phi(x_j,t) - \phi(x_i,t))$, with $\mathcal{N}_{x_i}$ the set of closest neighbors of $x_i$ in $\mathcal{S}$ (and $|\mathcal{N}_{x_i}|$ its cardinality). $\lambda$ denotes the diffusivity, $\phi(x_i,t)$ the population and $\nu(x_i,t)$ the resources on site $x_i$ at time $t$. Eq.\ref{PAM} is very well studied and appears under disguise in many physical contexts, such as random interface growth, directed polymers or turbulence \cite{PhysRevE.96.010102,HHZ,konig2016parabolic}. It mainly differs from the famous Anderson Hamiltonian (AH) by the temporal dependence of $\nu(x,t)$. The vast majority of the literature assumes in fact $\nu(x,t)$ to be white noise, but it will be crucial for us to fix an environmental timescale $\tau$. Therefore we set $\nu(x,t)$ as an Ornstein-Uhlenbeck (OU) process \cite{gueudre2014explore,PhysRevE.95.042134}:
\begin{align}
d \nu_{\tau}(x,t) = -\frac{\nu_{\tau}(x,t)}{\tau} dt + \frac{\sigma}{\tau} dW(x,t) \label{OU_scaling} \\
\langle \nu_{\tau}(x,t_1) \nu_{\tau}(y,t_2) \rangle = \frac{\sigma ^2}{2 \tau} e^{-\frac{|t_1-t_2|}{\tau}} \delta_{x,y}
\end{align}
with $W(x,t)$ a Wiener process. We fix $\sigma^2 = 1/2$ in the following. To probe the role of geometry, we consider various regular graphs $\mathcal{S}$ of degree $K$: the Euclidian grid $\mathcal{E}_d$ of dimension $d = K/2$ and the (locally tree-like) random regular graph $\mathcal{T}_K$ (with simulations for $d=1,2,3$, $K=3, 4$). The number of sites of the system is denoted $M$. Finally, the growth of the population $\phi(x_i,t)$ can be tracked through its asymptotic \textit{mean growth rate} $\bar{c}_{\tau}(\lambda)$ and the more relevant \textit{typical growth rate} $c_{\tau}(\lambda)$ (also called free energy):
\begin{align}
\bar{c}_{\tau}(\lambda) :=  \lim_{t \rightarrow \infty} \frac{\log \langle  \phi(x_i,t) \rangle}{t}  \text{   ,    } c_{\tau}(\lambda):=\lim_{t \rightarrow \infty}  \frac{\langle \log \phi(x_i,t) \rangle}{t} 
\end{align}
with $\langle \cdots \rangle$ denoting the average over the resources $\nu(x,t)$. In particular, $c_{\tau}(\lambda)$ is the most important observable, being the growth rate observed almost surely, for typical realizations of $\nu_{\tau}$.  By symmetry, those rates do not depend on the choice of $x_i$.

The PAM presents two interesting features. First, for any $\tau \geq 0$, it has a sharp localization transition $\lambda_c$, where the field $\phi(x,t)$ suddenly goes from a delocalized to a localized phase. Below a critical $\lambda_c$,  $c_{\tau}(\lambda) < \bar{c}_{\tau}(\lambda)$, while above, $c_{\tau}(\lambda) = \bar{c}_{\tau}(\lambda)$; the gap $\Delta c_{\tau}(\lambda) = \bar{c}_{\tau}(\lambda) - c_{\tau}(\lambda)$ grows with localization and reaches its largest value at $\lambda=0$. Second, it was found in \cite{gueudre2014explore} that $c_{\tau}(\lambda)$ exhibits a maximum at some $\lambda_m$ when $\tau$ is strictly positive, congruous with the exploration-exploitation paradigm. Therefore, this model presents both an optimal growth point at $\lambda_m$ and a risky, condensed phase below $\lambda_c$. In the following, we study the dependence of $c_{\tau}(\lambda)$, $\lambda_m$ and $\lambda_c$  with respect to $\tau$ and the structure of $\mathcal{S}$. The main tool of the analysis is the Feynman-Kac representation of the solution $\phi(x,t)$ of Eq.\ref{PAM} (see Appendix A):
\begin{align}
\phi(x,t) = E_{\pi_{X}} \left[ e^{\frac{1}{\sqrt{\lambda}} \int_0 ^t \nu_{\lambda \tau}(X(s),t-s) ds} \right] \label{FK}
\end{align}
$E_{\pi_{X}}[\cdotp]$ is the expectation over the measure $\pi_{X}$ of a continuous-time random walk $X$ on $\mathcal{S}$, with exponential jump distribution. Let us first investigate how the geometry of $\mathcal{S}$ affects $c_{\tau}(\lambda)$ and $\bar{c}_{\tau}(\lambda)$.

\begin{figure}[hbt!]
\includegraphics[width=0.48\textwidth]{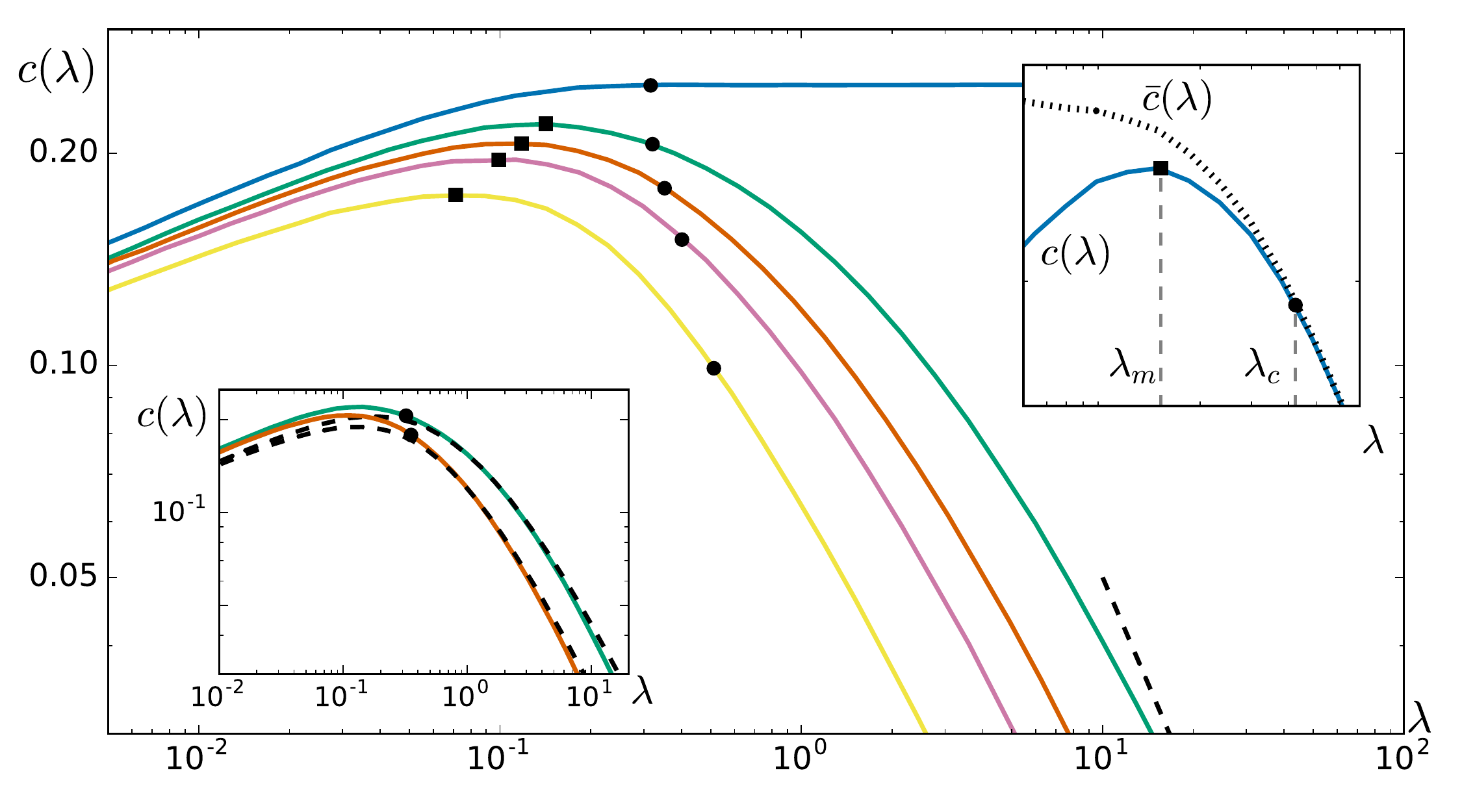} 
   \caption{Growth rates $c_{\tau}(\lambda)$ vs. $\lambda$ on the regular graph $\mathcal{T}_{K=4}$, for various values of $\tau = 0, 1/\sqrt{2},4/3,2$ and $5$ (from top to bottom). The dashed line is the asymptotics Eq.\ref{largelamb}. Circles mark the value of $\lambda_c$ on each curve, and squares mark $\lambda_m$. The white noise case (blue, top curve) reaches a plateau for $\lambda>\lambda_c$. $M=10^6$, $t_{tot}=100$. \textbf{(Top Inset)}: Pictorial representation of $c_{\tau}(\lambda)$ (in full) and $\bar{c}_{\tau}(\lambda)$ (in dashed), with the optimum $\lambda_m$ and the transition $\lambda_c$, for some $\tau>0$. \textbf{(Bottom Inset)}: Comparison of $c_{\tau}(\lambda)$ for both the grid $\mathcal{E}_{d=2}$ (in full) and $\mathcal{T}_{K=4}$ (in dashes), for $\tau=1/\sqrt{2},2$.}
   \label{growth_curves}
\end{figure}

\textit{Functional shapes of $c_{\tau}(\lambda)$ and $\bar{c}_{\tau}(\lambda)$}. The rates $c_{\tau}(\lambda)$ and $\bar{c}_{\tau}(\lambda)$ adopt similar shapes for all the analysed graphs. When $\tau>0$, they are very different at small $\lambda$, but merge above some $\lambda_c$ and decay to $0$ at large $\lambda$ (see Fig.\ref{growth_curves} for $\mathcal{S} = \mathcal{T}_{K=4}$ and Fig.\ref{growth_curves} top inset for a generic picture). This contrasts with the case $\tau=0$, where both rates reach a plateau of value $\sigma^2/2$ above $\lambda_c$. The analysis with no exploration is simple enough, leading to $c_{\tau}(\lambda= 0) = 0$ and $\bar{c}_{\tau}(\lambda= 0) = \sigma^2/2$, independently of $\mathcal{S}$ and $\tau$. We now turn onto the limits of small and large diffusivities. $c_{\tau}(\lambda)$ at small $\lambda$ can be estimated from a saddle point technique in the exponential term of Eq.\ref{FK} (see Appendix B), leading to the scaling form:
\begin{align}
c_{\tau}(\lambda) \propto \frac{\sigma^2}{4 \log(K/ \lambda)} \label{smalllamb}
\end{align}
(see \cite{carmona2001asymptotics} for a rigorous proof with $\mathcal{S}=\mathcal{E}_d$). The log dependence is quite striking, as it is not differentiable in $0$: typical growth is tremendously improved by a small amount of diffusion. Furthermore, the saddle point argument demonstrates that $c_{\tau}(\lambda)$ mainly depends on the degree of $\mathcal{S}$ for $\lambda \ll 1$. Hence $\mathcal{T}_{K = 2d}$ should be a rather good approximation of $\mathcal{E}_d$, as observed numerically in the bottom inset of Fig.\ref{growth_curves}. The reader may notice that Eq.\ref{FK} is akin to a partition function, each walk $X$ being a configuration of energy $e(X,t)=-\int_0^t \nu_{\lambda \tau}(X(s),t-s) ds$ and temperature $T=\sqrt{\lambda}$. Analogously to a free energy, $c_{\tau}(\lambda)$ is determined at small $\lambda$ by few low-energy configurations. Yet walks have more freedom to perform such optimization over well-connected graphs. It is therefore natural to expect $c_{\tau}(\lambda)$ to improve with the graph degree, given $\lambda$ is small enough. Indeed we checked numerically that, sufficiently close to $0$, $c_{\tau}(\lambda)$ increases with the dimension of $\mathcal{E}_d$ or the degree of $\mathcal{T}_K$.

Factors controlling the large $\lambda$ limit are quite different. The dominant term of the decay can be computed by a perturbative expansion of Eq.\ref{FK} (see Appendix C), leading to the result:
\begin{align}
c_{\tau}(\lambda) \sim \bar{c}_{\tau}(\lambda) \sim \frac{\sigma^{2}}{2\lambda \tau} G\left(0,0;1 \right) \quad \text{for} \quad \lambda \tau \gg 1 \label{largelamb}
\end{align}
with $G(x,y;z)$ the lattice Green function of the Laplacian on $\mathcal{S}$. Its analytical expression for $\mathcal{T}_K$ and $\mathcal{E}_d$ can be found in Appendix C. The decay of $c_{\tau}(\lambda)$ is linked to the way diffusion acts on $\mathcal{S}$.  It  more specifically depends on the transient or recurrent nature of the random walk $X$ over $\mathcal{S}$ \cite{doob2012classical}. In this respect, locally tree-like graphs and Euclidian grids are very different, as the absence of short loops in $\mathcal{T}_K$ greatly enhances the transience of walks, and decays on $\mathcal{E}_d$ and $\mathcal{T}_K$ coincide only in the limit $d = K = \infty$. More generally, the rates' decay is stronger for graphs with small return probabilities. Hence, while slowly diffusive population grow more easily on high-degree graphs, fast moving ones prefer both loopy or low dimensional spaces. 

\begin{figure}[hbt!]
\includegraphics[width=0.48\textwidth,trim={0 0 0 -1.0cm}]{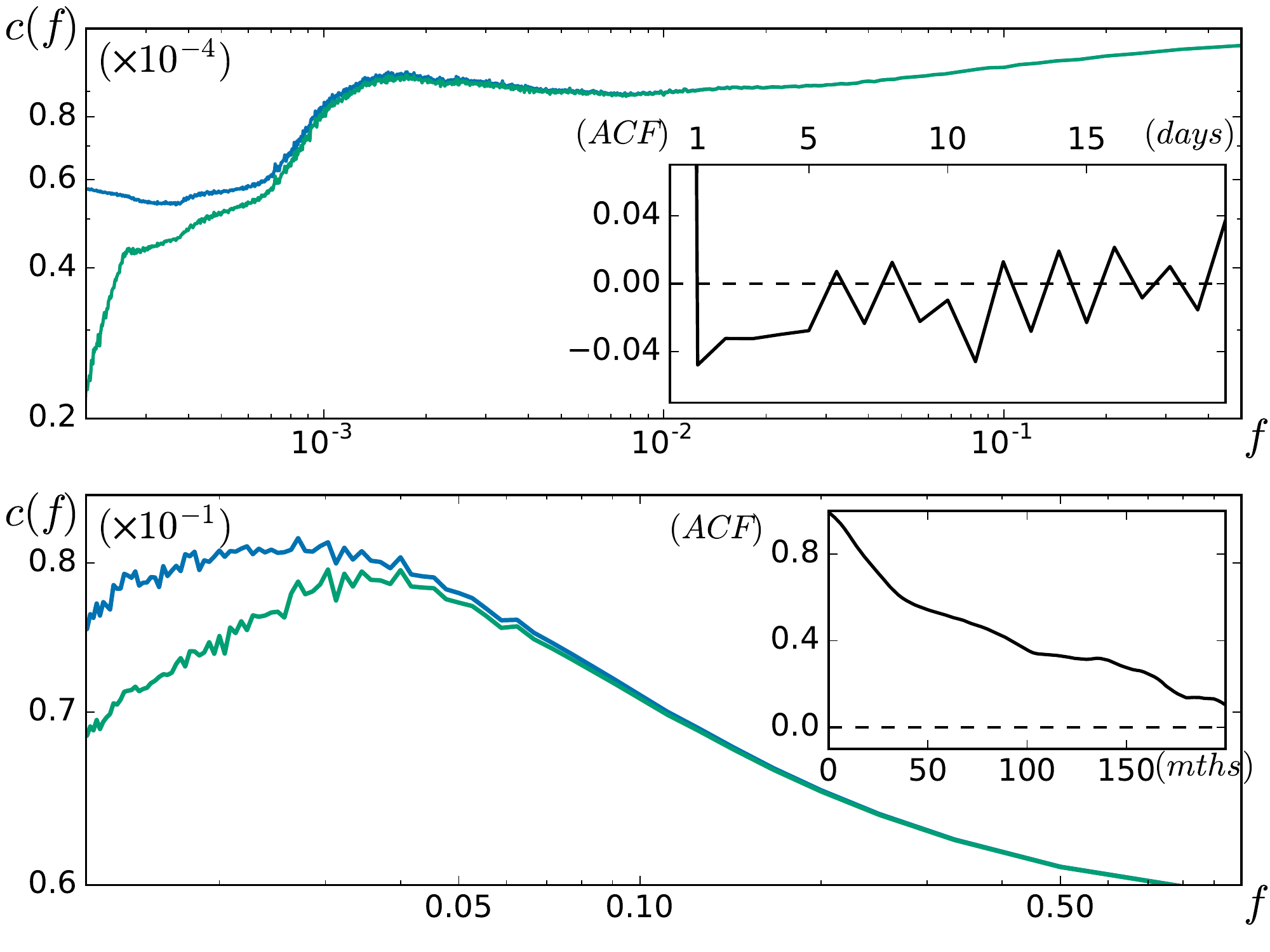} 
   \caption{\textbf{(Top)} Portfolio returns $\bar{c}(f)$ (upper curve) and $c(f)$ (lower curve) as a function of the rebalancing frequency $f$, for a $388$ NASDAQ stocks portfolio. (Inset) ACF of the data log-returns as a function of the lag in days. Negative correlations last up to $5$ days. \textbf{(Bottom)} Same returns, but for a portfolio of $23$ government bonds. (Inset) ACF of the data log-returns as a function of the lag in months. A correlation time of order $70$ months is observed.}
   \label{pos_hist}
\end{figure}

We can deduce that there is no overall better topology for growth. Remark however that, according to this analysis, growth curves over graphs with equal degree $K$ should be rather similar, a verdict supported by the inset of Fig.\ref{pos_hist} where we compare $c_{\tau}(\lambda)$ for $\mathcal{T}_{K=4}$ and $\mathcal{E}_{d=2}$. With the role of geometry clarified, we now examine the optimum and localization points in greater details. 

\begin{figure}[hbt!]
\includegraphics[width=0.49\textwidth,trim={0.5cm 0 0 0}]{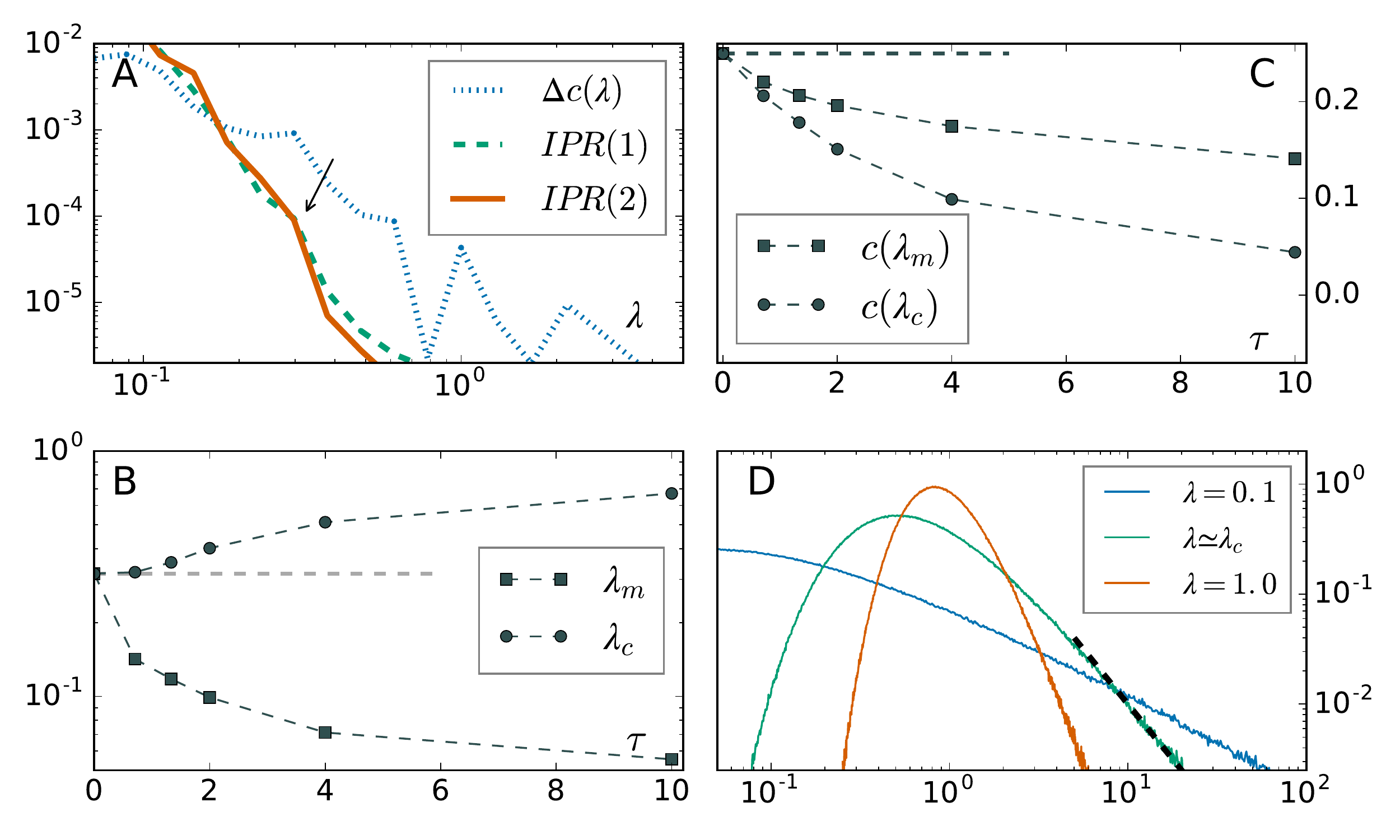} 
   \caption{Behaviour of the growth rate over $\mathcal{S}=\mathcal{T}_{K=4}$.\textbf{(A)} $\Delta c_{\tau}(\lambda) = \hat{c}_{\tau}(\lambda) - c_{\tau}(\lambda)$ and the partipation ratio $\text{IPR}(\lambda)$ for two systems sizes: $(1) \, M=10^6$ and $(2) \,M=2 \times 10^6$, and for $\tau=1.41$. The arrow highlights the inferred $\lambda_c \simeq 0.32$. \textbf{(B)} $\lambda_m$ and $\lambda_c$ as functions of $\tau$. The horizontal dashed line marks the common value $\lambda_{m/c}(\tau=0) \simeq 0.31$. \textbf{(C)} $c_{\tau}(\lambda_m)$ and $c_{\tau}(\lambda_c)$ as functions of $\tau$. The horizontal dashed line marks the common value $c_{\tau=0}(\lambda_{m/c}) = 1/4$ at $\tau=0$. \textbf{(D)} Histograms of the normalized population stationary distribution $P(\phi(x,t=\infty)/\langle \phi(x,t=\infty)\rangle)$ for $\lambda=0.1,0.41 (\simeq \lambda_c)$ and $1.0$ (from the broadest to the most narrow) and $\tau=2$. The dashed line is a power-law of exponent $-2$.}
   \label{lambda_anal}
\end{figure}

\textit{Behaviours of $\lambda_m$ and $\lambda_c$}. Balancing Eq.\ref{smalllamb} and Eq.\ref{largelamb} also provide an estimate of $\lambda_m$. But to pin down the optimal diffusion on any given graph, a more compelling argument stems from considering the diffusive timescale fixed by $\lambda$. Let us first assume there exists a resource-rich patch $\mathcal{P}$ in $\mathcal{S}$, that should last roughly for a duration $\tau$. The exit time $\tau_{ex}$ of the random walk $X$, starting at the center $x_0$ of this patch, obeys $E_{\pi_X}[\tau_{ex}] = \int _{\mathcal{P}} G_{\mathcal{P}}(x_0,y;1) dy$ \cite{doob2012classical,benichou2008exit}, where $G_{\mathcal{P}}(x,y;z)$ is again the diffusion Green function, but on $\mathcal{P}$ with Dirichlet boundary conditions. Equating both timescales gives $\lambda_m \sim \left( \tau \int _{\mathcal{P}} G_{\mathcal{P}}(x_0,y;1) dy \right)^{-1}$. While we do not state what the typical size of $\mathcal{P}$ is, this estimate shows that $\lambda_m$ decays as $1/\tau$, confirmed by numerics (see Fig.\ref{lambda_anal} (B)). Interestingly, given that random walks exit patches more easily on loop-free graphs, it also suggests that optimal growth will be found at larger $\lambda_m$ in clustered graphs (just as observed at the bottom Inset of Fig.\ref{growth_curves}). For regular graphs such as $\mathcal{E}_d$ or $\mathcal{T}_K$, we also recover the standard diffusive scaling: denoting $R$ the radius of the abundant patch, $\lambda_m \sim R^2/(K \tau)$. 

Unlike for $\lambda_m$, there exists a vast literature about the localization transition $\lambda_c$ \cite{HHZ,konig2016parabolic}. We recall that $\lambda_c$ is the value of the diffusivity below which $\Delta c_{\tau}(\lambda)= \bar{c}_{\tau}(\lambda) - c_{\tau}(\lambda) \neq 0$. Alas, $\lambda_c$ is notoriously difficult to estimate from $\Delta c_{\tau}(\lambda)$ alone. Another useful observable is the Inverse Participation Ratio $\text{IPR}(\lambda) = \left(\sum_{x_i \in \mathcal{S}} \phi(x_i,t)^2 \right)/\left(\sum_{x_i \in \mathcal{S}} \phi(x_i,t)\right)^2$, also called the Simpsons index in ecology \cite{simpson1949measurement}, or the Herfindahl index in economy \cite{federal2014department}. It goes to $0$ as $M^{-1}$ in the delocalized phase and remains of order $1$ otherwise, giving a very close upper bound of $\lambda_c$ (see Appendix D, and also \cite{monthus2006numerical} for more details). A non-vanishing IPR relates to a diverging mean of the stationary distribution of the field $P(\phi(.,t))$. We tracked both $\Delta c_{\tau}(\lambda)$ and the IPR for different size systems in order to estimate $\lambda_c$ (see Fig.\ref{lambda_anal} A and B). Sharp localization transitions provably exist on $\mathcal{E}_{d \geq 3}$ and $\mathcal{T}_{K \geq 3}$. A simple coarse-graining argument suggests that $\lambda_c \sim \tau$ at large $\tau$. Intuitively, slow fluctuations favor the localized phase, as long-lived abundent patches allow for concentrated growth to be efficient. Therefore, $\lambda_m$ and $\lambda_c$ are controlled by contrastive mechanisms: the first goes to $0$ while the second diverges as the environmental timescale increases. Fig.\ref{lambda_anal} (B) illustrates this growing gap. Note also that, in this model, diffusion is overall a less and less efficient strategy as $\tau$ increases: $c_{\tau}(\lambda_m)$ and $c_{\tau}(\lambda_c)$ go to $0$ with $\tau \rightarrow \infty$ (see Fig.\ref{lambda_anal} (C)).

Hence for large $\tau$, the optimum is deep in the risky condensed phase $\lambda_m \ll \lambda_c$. It is actually always true: again from Eq.\ref{FK}, one obtains that $\bar{c}_{\tau}(\lambda)$ is monotonically decreasing with $\lambda$, starting from $\hat{c}_{\tau}(\lambda=0)=\sigma^2/2$. Given that $c_{\tau}(\lambda)= \bar{c}_{\tau}(\lambda)$ above $\lambda_c$, this simple fact implies that $\lambda_m < \lambda_c$ at any $\tau$. One of our main conclusions, that optimal growth is \textit{always localized}, has several interesting repercussions. Firstly, recall that in the low $\lambda$ regime, diffusive growth favours high-degree graphs. We expect it to remain true for $c_{\tau}(\lambda_m)$, especially when the optimum lies deeply in the localized phase for large $\tau$. Secondly, anti-trust laws \cite{federal2014department} or ecosystems monitoring \cite{simpson1949measurement}, forcing the IPR to $0$, usually fight against the tendency to localize. Systems optimizing their own growth under such restrictions would marginally end up at the transition $\lambda_c$. In this zone, the population distribution develops heavy tails of exponent around $-2$ (see Fig.\ref{lambda_anal} D), a range typically observed in the \textit{Zipf law} literature, where the above mechanism may be at work \cite{bouchaud2000wealth,fiaschi2010economic}.

Let us summarize the important elements drawn from the analysis of Eq.\ref{PAM}. Growth depends on two times scales set by $\tau$ and $\lambda$ and on the underlying geometry of $\mathcal{S}$, best understood through the Green function of the Laplacian operator. Networks over which growth strives have different features at low and high diffusivities: slow populations favor large degrees, fast ones prefer recurrent spaces. Finally, $\tau$ slaves $\lambda_c \sim \tau$ and $\lambda_m \sim 1/\tau$ in an antipodal way, so much that the optimal point always lies in the localized phase, except for $\tau=0$. Hence, we expect very connected spaces to provide the greatest overall returns $c_{\tau}(\lambda_m)$, but at the cost of a risky population distribution with heavy tails.

\textit{Empirical study with financial data}. What is the practical value of the above conclusions ? They were derived for a specific stochastic noise $\nu_{\tau}(x,t)$ and diffusive dynamics. Yet they are consistent with general principles, such as the exploration-exploitation trade-off, and should exhibit some universality. As a check, we apply them to the problem of portfolio optimization, where funds invest money over a set of assets (the nodes of the graph $\mathcal{S}$). Each invested amount grows exponentially according to a fluctuating log-return, the analog of $\nu_{\tau}(x,t)$. It has been long known that investments left alone tend to localize on few well-performing assets, echoing the localization detailed above. To avoid these risky positions, funds employ rebalancing strategies, of which we adopt here a rudimentary and common embodiment, the calendar rebalancing: after a fixed temporal interval $T_b$, the total money is equally reinvested over all the neighbors assets (usually the whole portfolio) \footnote{Note that we neglect many factors such than reinvestmnet costs, as they are irrelevant for our discourse.}.

For the empirical data, we consider the S\&P listed stocks, starting from January 1998 to December 2013, with daily open and close ticks \cite{nasdaq}. We pick the set of $M=388$ companies with the longest history and simulate a portfolio with a small initial investment on each, applying a calendar rebalancing of frequency $f = 1/T_b$. We then monitor both the typical and mean returns $c(f)$ and $\bar{c}(f)$ of our invesment as a function of $f$, after $10$ years (see Appendix E). The results are presented in Fig.\ref{pos_hist} Top, where $c(f)$ surprisingly shows no optimum at all. In fact, it keeps growing as $f \rightarrow \infty$. Yet it is consistent with the financial wisdom that equities' returns are far from uncorrelated gaussian noises. The assets' returns are even temporally \textit{anti}-correlated (as show by the auto-correlation function (ACF) presented in the inset of Fig.\ref{pos_hist} Top). Because there is no incentive to spend any time on a given asset, rebalancing should be as frequent as possible $f_{m} = \infty$, in agreement with our analysis. 

Assets with correlated returns do exist, for example in government bonds trading. Governments around the world emit debt bonds, and their interest rate is computed every month according to the economical performance of the emitting state. Bond owners can speculate over these interest rates, that usually follow the slower timescale set up by their countries. We consider a dataset of $23$ countries over a total of $285$ months \cite{govbonds}, and apply the same type of rebalancing (see Fig.\ref{pos_hist} Bottom). This time, a clear optimum is observed, at roughly $f=0.02 \text{   mth}^{-1}$. In the inset of Fig.\ref{pos_hist} Bottom, we show again the ACF and find clear positive correlations over a timescale of $50$ months, consistent with the relation $f_m \simeq 1/\tau$. The optimal strategy creates a localized portfolio, with $\bar{c}(f_m)$ somewhat higher than $c(f_m)$. Instead of redistributing the wealth over all assets, we also tried a more local redistribution over $\mathcal{S}=\mathcal{T}_{K=4}$ (see Appendix E). The results are completely similar, but with a smaller optimal return $c(f_m)$ and a slower decay at large $f$. Hence, even if the returns are not Gaussian and the dynamics not diffusive, those findings parallel our analysis that, in a localized phase, very connected graphs provide a better bed for growth, and we conjecture that the complete graph, out of all, leads to the largest optimal return $c(f_m)$. Less connected graphs may be preferred for faster and safer rebalancing. Remarkably, we did not perform any kind of rescaling on the data, except to remove the base trend of the market. The devised principles are impressively robust with respect to heterogeneities in the stochastic environment. 
   
Identical conclusions could be reached for other time-correlated products such as trading pairs or similar processes in different contexts such as world trading of basic goods \cite{caraglio2016export} and fitness landscapes of microbial populations \cite{visco2010switching}. We believe our conclusions are robust enough to provide a general understanding of multiplicative growth coupled with exploration. This study can be furthered in many directions. Suprisingly, a detailed study of Eq.\ref{PAM} on the tree-like lattices $\mathcal{T}_{K=2d}$, although a good approximation to the harder Euclidian case, is still lacking and warrants a more in-depth look (\cite{tarquini2017critical} reached similar conclusions for the related Anderson Hamiltonian). We also skimmed over the microscopic details of the population distribution, focusing on the macroscopic growth rates. Hinsights gained from Eq.\ref{PAM} about how population spread already have a fecond history \cite{bouchaud2000wealth} and we leave its study in the present case for latter work. Last but not least, we ignored heterogeneity. The spaces studied here are regular graph, variances and correlation times of the noise are chosen equal for all sites. Although heterogeneity does not invalidate our conclusions, as shown by the empirical study, its role requires a more thorough inquiry. 


\bibliographystyle{unsrt}
\bibliography{optimal_principles}

\begin{thebibliography}{10}

\bibitem{royama2012analytical}
Tomoo Royama.
\newblock {\em Analytical population dynamics}, volume~10.
\newblock Springer Science \& Business Media, 2012.

\bibitem{bouchaud2003theory}
Jean-Philippe Bouchaud and Marc Potters.
\newblock {\em Theory of financial risk and derivative pricing: from
  statistical physics to risk management}.
\newblock Cambridge university press, 2003.

\bibitem{PhysRevLett.119.140603}
Andrea Falc\'on-Cort\'es, Denis Boyer, Luca Giuggioli, and Satya~N. Majumdar.
\newblock Localization transition induced by learning in random searches.
\newblock {\em Phys. Rev. Lett.}, 119:140603, Oct 2017.

\bibitem{PhysRevLett.113.238101}
O.~B\'enichou and S.~Redner.
\newblock Depletion-controlled starvation of a diffusing forager.
\newblock {\em Phys. Rev. Lett.}, 113:238101, Dec 2014.

\bibitem{PhysRevLett.108.088103}
Vincent Tejedor, Raphael Voituriez, and Olivier B\'enichou.
\newblock Optimizing persistent random searches.
\newblock {\em Phys. Rev. Lett.}, 108:088103, Feb 2012.

\bibitem{benichou2017optimally}
O~Benichou, U~Bhat, PL~Krapivsky, and S~Redner.
\newblock Optimally frugal foraging.
\newblock {\em arXiv preprint arXiv:1711.03610}, 2017.

\bibitem{diaz2016global}
Sandra D{\'\i}az, Jens Kattge, Johannes~HC Cornelissen, Ian~J Wright, Sandra
  Lavorel, St{\'e}phane Dray, Bj{\"o}rn Reu, Michael Kleyer, Christian Wirth,
  I~Colin Prentice, et~al.
\newblock The global spectrum of plant form and function.
\newblock {\em Nature}, 529(7585):167--171, 2016.

\bibitem{ibarra2002escherichia}
Rafael~U Ibarra, Jeremy~S Edwards, and Bernhard~O Palsson.
\newblock Escherichia coli k-12 undergoes adaptive evolution to achieve in
  silico predicted optimal growth.
\newblock {\em Nature}, 420(6912):186--189, 2002.

\bibitem{caraglio2016export}
Michele Caraglio, Fulvio Baldovin, and Attilio~L Stella.
\newblock Export dynamics as an optimal growth problem in the network of global
  economy.
\newblock {\em Scientific reports}, 6, 2016.

\bibitem{gärtner2010}
J.~Gartner, F.~den Hollander, and G.~Maillard.
\newblock Intermittency on catalysts: Voter model.
\newblock {\em Ann. Probab.}, 38(5):2066--2102, 09 2010.

\bibitem{korolev2014turning}
Kirill~S Korolev, Joao~B Xavier, and Jeff Gore.
\newblock Turning ecology and evolution against cancer.
\newblock {\em Nature Reviews Cancer}, 14(5):371--380, 2014.

\bibitem{simpson1949measurement}
Edward~H Simpson.
\newblock Measurement of diversity.
\newblock {\em Nature}, 1949.

\bibitem{federal2014department}
Federal~Trade Commission et~al.
\newblock Department of justice. horizontal merger guidelines. august 19, 2010,
  2014.

\bibitem{Gatenby4894}
Robert~A. Gatenby, Ariosto~S. Silva, Robert~J. Gillies, and B.~Roy Frieden.
\newblock Adaptive therapy.
\newblock {\em Cancer Research}, 69(11):4894--4903, 2009.

\bibitem{PhysRevE.96.010102}
Pierre Le~Doussal and Thimoth\'ee Thiery.
\newblock Diffusion in time-dependent random media and the kardar-parisi-zhang
  equation.
\newblock {\em Phys. Rev. E}, 96:010102, Jul 2017.

\bibitem{HHZ}
T.~Halpin-Healy and Y.-C. Zhang.
\newblock Kinetic roughening phenomena, stochastic growth, directed polymers
  and all that. aspects of multidisciplinary statistical mechanics.
\newblock {\em Physics Reports}, 254(4–6):215--414, 1995.

\bibitem{konig2016parabolic}
Wolfgang K{\"o}nig.
\newblock {\em The Parabolic Anderson Model: Random Walk in Random Potential}.
\newblock Birkh{\"a}user, 2016.

\bibitem{gueudre2014explore}
Thomas Gueudr{\'e}, Alexander Dobrinevski, and Jean-Philippe Bouchaud.
\newblock Explore or exploit? a generic model and an exactly solvable case.
\newblock {\em Physical review letters}, 112(5):050602, 2014.

\bibitem{PhysRevE.95.042134}
Thomas Gueudr\'e.
\newblock Localized growth and branching random walks with time correlations.
\newblock {\em Phys. Rev. E}, 95:042134, Apr 2017.

\bibitem{carmona2001asymptotics}
Ren{\'e} Carmona, Leonid Koralov, and Stanislav Molchanov.
\newblock Asymptotics for the almost sure lyapunov exponent for the solution of
  the parabolic anderson problem.
\newblock {\em Random Operators and Stochastic Equations}, 9(1):77--86, 2001.

\bibitem{doob2012classical}
Joseph~L Doob.
\newblock {\em Classical potential theory and its probabilistic counterpart:
  Advanced problems}, volume 262.
\newblock Springer Science \& Business Media, 2012.

\bibitem{benichou2008exit}
Olivier B{\'e}nichou and Jean Desbois.
\newblock Exit and occupation times for brownian motion on graphs with general
  drift and diffusion constant.
\newblock {\em Journal of Physics A: Mathematical and Theoretical},
  42(1):015004, 2008.

\bibitem{monthus2006numerical}
C{\'e}cile Monthus and Thomas Garel.
\newblock Numerical study of the directed polymer in a 1+ 3 dimensional random
  medium.
\newblock {\em The European Physical Journal B-Condensed Matter and Complex
  Systems}, 53(1):39--45, 2006.

\bibitem{bouchaud2000wealth}
Jean-Philippe Bouchaud and Marc M{\'e}zard.
\newblock Wealth condensation in a simple model of economy.
\newblock {\em Physica A: Statistical Mechanics and its Applications},
  282(3):536--545, 2000.

\bibitem{fiaschi2010economic}
Davide Fiaschi and Matteo Marsili.
\newblock Economic interactions and the distribution of wealth.
\newblock {\em Econophysics and Economics of Games, Social Choices and
  Quantitative Techniques}, pages 61--70, 2010.

\bibitem{nasdaq}
QuantQuote.
\newblock Nasdaq stocks daily ticks.
\newblock \url{https://quantquote.com/historical-stock-data}.
\newblock Accessed: 2017-10-02.

\bibitem{govbonds}
OECD.
\newblock Government bonds semi-yearly ticks.
\newblock
  \url{https://data.oecd.org/interest/long-term-interest-rates.htm\#indicator-chart}.
\newblock Accessed: 2017-10-10.

\bibitem{visco2010switching}
Paolo Visco, Rosalind~J Allen, Satya~N Majumdar, and Martin~R Evans.
\newblock Switching and growth for microbial populations in catastrophic
  responsive environments.
\newblock {\em Biophysical journal}, 98(7):1099--1108, 2010.

\bibitem{tarquini2017critical}
Elena Tarquini, Giulio Biroli, and Marco Tarzia.
\newblock Critical properties of the anderson localization transition and the
  high-dimensional limit.
\newblock {\em Physical Review B}, 95(9):094204, 2017.

\bibitem{monthus_texier}
Cécile Monthus and Christophe Texier.
\newblock Random walk on the bethe lattice and hyperbolic brownian motion.
\newblock {\em Journal of Physics A: Mathematical and General}, 29(10):2399,
  1996.

\bibitem{PhysRevE.54.2084}
Daniel~T. Gillespie.
\newblock Exact numerical simulation of the ornstein-uhlenbeck process and its
  integral.
\newblock {\em Phys. Rev. E}, 54:2084--2091, Aug 1996.

\end{thebibliography}

\clearpage
\onecolumngrid
\appendix

\section{The Feynman-Kac formula for the PAM}

In this section, we recall the Feynman-Kac formula (or path integral) for Eq.\ref{PAM}. The Feynman-Kac formalism reduces the solution of a partial differential equation to an expectation value over some random process. It is usually formulated with terminal conditions, so one needs first to revert the time of Eq.\ref{PAM} by the change $s \rightarrow t-s$, leading to:
\begin{align}
\phi(x_i,t) &= E_{\pi_{X_{\lambda}}}\left[ \exp \left(\int_0 ^t \nu_{\tau}(X(s),t-s) ds \right) \right] \label{FK_appendix} \\
\phi(x_i,0) &= 1 \quad \forall x_i \in \mathcal{S}
\end{align}
The expectation $E_{\pi_{X_{\lambda}}}$ is taken over the distribution of continuous time random walks $X_{\lambda}$ in $\mathcal{S}$, that wait a random time sampled from an exponential distribution of parameter $\lambda$, before hopping onto neighbours, with initial conditions $X_{\lambda}(0)=x_i$.
 
Note that with we are considering $\nu_{\tau}(x,.)$ as an Ornstein-Ulhenbeck process, which bypasses some difficulties in defining the integral Eq.\ref{FK_appendix}. We also use the fact that, if its initial value is draw from its stationary distribution, $\nu_{\tau}(x,.)$ is time reversible, so that we can equivalently consider the slightly simpler:
\begin{align}
\phi(x_i,t) &= E_{\pi_{X_{\lambda}}}\left[\exp \left(\int_0 ^t \nu_{\tau}(X(s),s) ds \right)\right] \label{FK2_appendix}
\end{align}

The clearest way to derive asymptotics is to introduce sets of rescaled variables.  Although we always present the final results in the original unscaled variables, the $\lambda$ dependence in the distribution $\pi_{X_\lambda}$ can be absorbed by choosing $t \rightarrow \lambda t$. With this rescaling, Eq.\ref{PAM} reads:
\begin{align}
\partial_{t} \phi(x,t) &= \Delta_x \phi(x,t) + \frac{1}{\sqrt{\lambda}}\nu_{1}(x,t) \phi(x,t) \\
d  \nu_{\tau \lambda}(x,t) &= -  \frac{\nu_{\tau \lambda}(x,t)}{\lambda \tau} dt + \frac{\sigma}{\lambda \tau} dW(x,t)
\end{align}
The rescaled noise $\nu_{\tau \lambda}$ has now correlations:
\begin{align}
\langle \nu_{\tau \lambda}(x,t)\nu_{\tau \lambda}(y,0) \rangle = \frac{\sigma^{2}}{2\tau\lambda}e^{-\frac{t}{\tau\lambda}}\delta (x-y) 
\end{align}
And the Feynman-Kac formula adopts the form presented in the main text Eq.\ref{FK} because of its analogy with a partition function:
\begin{align}
\phi(x,t) = E_{\pi_X} \left[\exp \left( \frac{1}{\sqrt{\lambda}} \int_0 ^t \nu_{\tau \lambda}(X(s),t-s) ds \right) \right]
\label{FK3_appendix}
\end{align}
with $ E_{\pi_X}$ being now the expectation over random walks with exponential time jump distribution of parameter $1$.

\section{The small diffusion behaviour} 
In this section we investigate the small $\lambda$ behaviour of $c_{\tau}(\lambda)$. Let us first consider the $\lambda=0$ case. Eq.\ref{PAM} reduces to:
\begin{align}
\partial_t \phi(x,t) = \nu_{\tau}(x,t)\phi(x,t)
\end{align}
which leads to the exponential of the OU process:
\begin{align}
\phi(x,t) = \exp \left(\int_0 ^t \nu_{\tau}(x,u) du \right)
\end{align}
In this limit, the difference between the quenched and annealed values is at its largest. Indeed:
\begin{align}
c_{\tau}(0) &= \lim_{t \rightarrow \infty} \frac{1}{t} \langle \log \phi(0,t) \rangle = \lim_{t \rightarrow \infty} \frac{1}{t} \langle \int_0 ^{\infty} \nu_{\tau}(0,u) du \rangle = 0
\end{align}
On the other hand:
\begin{align}
\lim_{t \rightarrow \infty} \frac{1}{t} \log  \langle  \phi(0,t) \rangle &= \lim_{t \rightarrow \infty} \frac{\tau \sigma^2}{2 t} \left(e^{-t/\tau} -1 +\frac{t}{\tau} \right) = \frac{\sigma^2}{2} 
\end{align}
This quantity is independent of $\tau$, which is the reason behing the original scaling of Eq.\ref{OU_scaling}: the accumulation of either white noise or Ornstein-Uhlenbeck increments are basically equivalent at large times.

The expansion at small $\lambda$ is best obtained with the set of variables of Eq.\ref{FK2_appendix}, that we recall:
\begin{align}
\phi(x_i,t) &= E_{\pi_{X_{\lambda}}} \left[\exp \left(\int_0 ^t \nu_{\tau}(X(s),s) ds \right) \right] \label{FKeq} \\
\phi(x_i,t=0) &= 1 \quad  \forall \, x_i \in \mathcal{S}
\end{align}
A path of the walk $X_{\lambda}$ is defined by the number of jumps $N$ up to time $t$, the list of the positions visited $\{X_{1},X_{2},..,X_{N}\}$ and the list of times at which those jumps occured $\{t_{1},t_{2},..,t_{N}\}$. We denote $D_N$ the set of all possible positions $N$-lists, with $|D_N|=K^N$, $K$ the degree of $\mathcal{S}$. Similarly, we denote $L_N$ the set of $N$ ordered possible times, up to $t$, drawn from $[0,t]$ with a density $\frac{N!}{t^N}$. Finally, the probability for $X_{\lambda}$ to perform $N$ jumps in $[0,t]$ follows a Poisson distribution of rate $\lambda$:
\begin{align}
P(N) = e^{-\lambda t} \frac{(\lambda t)^N}{N!}
\label{poisson_rate}
\end{align}
The expectation in Eq.\ref{FKeq} can thus be rewritten as:
\begin{align}
\phi(u,t) = \sum_{N=1}^{\infty} P(N)\sum_{\lbrace x_1,x_2...\rbrace \in D_{N}} \frac{1}{K^{N}} \int_{\lbrace t_1,t_2...\rbrace \in L_{N}}\frac{N!}{t^{N}} \exp \left(\int_0 ^{t_1} \nu_{\tau}(x_1,s) ds + \int_{t_1} ^{t_2} \nu_{\tau}(x_2,s) ds +... \right) dt_1 ... dt_N \label{FKeqFinal}
\end{align}
Now, we note that at small $\lambda$ (more exactly for $\lambda \tau \ll 1$), the average jump time is much larger than the correlation time of $\nu_{\tau}(x,.)$. Then the integral in Eq.\ref{FKeqFinal} can be split in approximately independent pieces, each corresponding to a sojourn period on one site of the walk. Assuming that the interval between jumps is always the same, and therefore roughly equal to $t/N$, we can determine the typical value of each integral splits:
\begin{align}
\langle \left( \int_{0}^{\frac{t}{N}} \nu_{\tau}(x,s) ds \right)^2\rangle & \simeq \frac{\sigma^{2} t}{N} \\
 \sum_{N}\int_{0}^{\frac{t}{N}} \nu_{\tau}(X(s),s) ds &\sim \sigma \sqrt{tN} \label{std_value}
\end{align}
We are interested in the behaviour of $\log(\phi(u,t))$. Plugging Eq.\ref{poisson_rate} and Eq.\ref{std_value} in Eq.\ref{FKeqFinal} and maximizing the argument of the exponential, we obtain a set of saddle point equations for the optimal number of jumps $N^*$. More precisely, the optimal temporal jumps density $\hat{N} = \frac{N^*}{t}$ follows:
\begin{align}
0 &= \log \left( \frac{\lambda}{K} \right) + \frac{\sigma}{2 \sqrt{\hat{N}}} \\
c_{\tau}(\lambda) &= -\lambda  + \hat{N} \log \left( \frac{\lambda}{K} \right) + \sigma \sqrt{\hat{N}}
\end{align}
Expanding for $\lambda \tau \ll 1$ leads to:
\begin{align}
c_{\tau}(\lambda) \propto \frac{\sigma^{2}}{4\log(\frac{K} {\lambda})} \label{smallL}
\end{align} 
up to a prefactor that depends on the nature of the disorder (rigorous results for the Euclidian case can be found in \cite{carmona2001asymptotics}). This derivation highlights some interesting facts: first note that $c_{\tau}(\lambda)$ is not differentiable in $0$, and has a logarithmic singularity, quite difficult to observe numerically. Moreover, because of the large sojourn time, the details of the noise $\nu(x,t)$ have little importance, as the central limit theorem takes place in Eq.\ref{std_value}, and this result extends to a more general class of stochastic processes. Also, the dependence in $\tau$ drops out, as long as $\lambda \tau \ll 1$.

Finally, we remark that Eq.\ref{smallL} provides a poor estimate that even becomes negative for $\lambda > 1/K$. The saddle points equations do not suffer from such problem, but we observe numerically that the approximation becomes unreliable even at moderate values of $\lambda$.

\section{The large diffusion behaviour}
In this section, we derive the decay of $c_{\tau}(\lambda)$ at large $\lambda$. We recall the rescaled form of the Feynman-Kac formula Eq.\ref{FK3_appendix}:
\begin{align}
\phi(x,t) = E_{\pi_X} \left[\exp \left( \frac{1}{\sqrt{\lambda}} \int_0 ^t \nu_{\tau \lambda}(X(s),s) ds \right) \right]
\label{rescaledFK}
\end{align}
In the large $\lambda$ regime, we expect $\bar{c}_{\tau}(\lambda)$ and $c_{\tau}(\lambda)$ to be equal, or at least very close, because the population is delocalized. Hence we equivalently look at $\bar{c}_{\tau}(\lambda)$, an quantity easier to compute. Averaging Eq.\ref{rescaledFK} over the disorder gives:
\begin{align}
\langle \phi(x,t) \rangle &= E_{\pi_X} \left[\langle \exp \left( \frac{1}{\sqrt{\lambda}}\int_0 ^t \nu_{\tau \lambda}(X(s),s) ds \right) \rangle\right] \\
\langle \phi(x,t) \rangle &= E_{\pi_X} \left[\exp \left( \frac{\sigma^{2}}{4\lambda^{2}\tau}\int_0 ^t \int_0 ^t \delta(X(t_{1})-X(t_{2}))e^{-\frac{|t_{1}-t_{2}|}{\tau\lambda}} dt_{1}dt_{2} \right) \right] \\
\langle \phi(x,t) \rangle &\sim E_{\pi_X} \left[1+\frac{\sigma^{2}}{4\lambda^{2}\tau}\left(\int_0 ^t \int_0 ^t \delta(X(t_{1})-X(t_{2}))e^{-\frac{|t_{1}-t_{2}|}{\tau\lambda}} dt_{1}dt_{2} \right) \right] \label{taylor_phi}
\end{align}
where in the last line, we used the hypothesis $\lambda \tau \gg 1$. We have to evaluate the following term:
\begin{align}
A=E_{\pi_X} \left[\frac{\sigma^{2}}{4\lambda^{2}\tau}\left(\int_0 ^t \int_0 ^t \delta(X(t_{1})-X(t_{2}))e^{-\frac{|t_{1}-t_{2}|}{\tau\lambda}} dt_{1}dt_{2} \right) \right]
\end{align}
Noting the parity of the integrand, we have:
\begin{align}
A&=E_{\pi_X} \left[\frac{\sigma^{2}}{2\lambda^{2}\tau}\left(\int_0 ^t \int_{0}^{t'} \delta(X(t'+u)-X(t'))e^{-\frac{u}{\tau\lambda}} du \,dt' \right) \right] \\
&=\frac{\sigma^{2}}{2\lambda^{2}\tau}\left(\int_0 ^t \int_{0}^{t'} P(0,0;u)e^{-\frac{u}{\tau\lambda}} du \, dt' \right)  \\
&=\frac{\sigma^{2}}{2\lambda^{2}\tau}\int_0 ^t P(0,0;u)e^{-\frac{u}{\tau\lambda}}(t-u) du 
\end{align}
$E_{\pi_X} [\delta(X(t'+u)-X(t'))]$ is the probability that $X$ comes back to its position at time t', after a delay $u$. Given the time translation invariance of $X$, and the space symmetry of the graphs considered in the paper, it simply is $P(0,0;u)$, the probability for $X$ to return to its origin after such a delay $u$. Therefore, from Eq.\ref{taylor_phi}:
\begin{align}
\log (\langle \phi(0,t) \rangle) &\sim \frac{\sigma^{2}}{2\lambda^{2}\tau}\int_0 ^t P(0,0;u)e^{-\frac{u}{\tau\lambda}}(t-u) du \\
\lim_{t \rightarrow \infty}\frac{\log (\langle \phi(0,t) \rangle)}{t} &\sim \frac{\sigma^{2}}{2\lambda^{2}\tau}\int_0 ^\infty P(0,0;u)e^{-\frac{u}{\tau\lambda}} du 
\label{HighLambda}
\end{align}
What can be said about this integral ? $P(0,0;u)$ is a well-known object that we can express as a sum over the number of jumps $N$ occuring between $0$ and $u$. If we note $P_{N}(0,0)$ the probability to come back at the origin after $N$ jumps, and $P(N)$ the probability (Eq.\ref{poisson_rate}) that $N$ jumps occured up to time $u$, we have :
\begin{align}
P(0,0;u)=\sum_{N=0}^{\infty} P_{N}(0,0)P(N) \\
P(0,0;u)=\sum_{N=0}^{\infty} P_{N}(0,0)e^{-u} \frac{u^N}{N!}
\end{align}
Injecting this expression back into Eq.\ref{HighLambda}, we get:
\begin{align}
c_{\tau}(\lambda) &\sim \frac{\sigma^{2}}{2\lambda^{2}\tau}\sum_{N=0}^{\infty}P_{N}(0,0)\int_0 ^\infty e^{- u} \frac{ u^N}{N!}e^{-\frac{u}{\tau\lambda}} du \\
&\sim \frac{\sigma^{2}}{2\lambda^{2}\tau}\frac{1}{1+\frac{1}{\tau\lambda}}\sum_{N=0}^{\infty}P_{N}(0,0)\left(\frac{1}{1+\frac{1}{\lambda\tau}}\right)^{N} \\
 &\sim \frac{\sigma^{2}}{2\lambda^{2}\tau}\frac{1}{1+\frac{1}{\tau\lambda}}G\left(0,0;\frac{1}{1+\frac{1}{\tau\lambda}}\right)
\end{align}
$G(0,0;z)$ is the generating function of return probabilities to the origin (also called the lattice Green function) $G(x,y;z)=\sum_{N}P_{N}(x,y)z^{N}$, a classical object for the study of Markov chains on lattices \cite{doob2012classical}. It has well-known expression for a variety of graphs, amongst which the Euclidian ones ($BZ$ is the Brilloin Zone). For the random regular graphs $\mathcal{T}_K$, we approximate the Green function by its expression over an infinite $d$-tree, using the fact that they have a common local limit in the large $N$ limit \cite{monthus_texier}:
\begin{align}
G(0,0;z)&=\frac{1}{(2\pi)^{d}}\int_{BZ}d^{d}k \frac{1}{1-z/(2d)\sum_{j \in \mathcal{N}_j} e^{ik.j}} \quad \text{for}\quad \mathcal{S}=\mathcal{E}_d \\
G(0,0;z) &\simeq 2 \frac{\frac{K-1}{K}}{\frac{K-2}{K}+\sqrt{1-\frac{4}{K^2}(K-1)z^2}} \quad \text{for}\quad \mathcal{S}=\mathcal{T}_K
\end{align}
Finally, we take again the limit of $\lambda \tau \gg 1$. $G(0,0;z)$ for $z$ close to $1$ is known to diverge on $\mathcal{E}_{d \leq 2}$ (as random walks over them are recurrent), while it does converge for $\mathcal{E}_{d \geq 3}$ and $\mathcal{T}_K$. More precisely:
\begin{align}
G(0,0;1-\epsilon) &\sim \frac{1}{\sqrt{2\epsilon}} \quad \text{for } \quad \mathcal{E}_{d=1} \\
G(0,0;1-\epsilon) &\sim \frac{1}{4\pi}\ln\left(\frac{1}{\epsilon}\right) \quad \text{for } \quad \mathcal{E}_{d=2}\\
G(0,0;1) &= \frac{\sqrt{6}}{32\pi^{3}}\Gamma\left(\frac{1}{24}\right)\Gamma\left(\frac{5}{24}\right)\Gamma\left(\frac{7}{24}\right)\Gamma\left(\frac{11}{24}\right) \approx 1.516 \quad \text{for } \quad \mathcal{E}_{d=3}\\
G(0,0;1) &\simeq \frac{K-1}{K-2} \quad \text{for } \quad \mathcal{T}_{K}
\end{align}
Taking into account these behaviours, and going back to the original variables (with $c \rightarrow c \lambda$), we retrieve the following results for $\lambda \tau \gg 1$ :
\begin{align}
c_{\tau}(\lambda) &\simeq \frac{\sigma^{2}}{2\sqrt{2\lambda\tau}} \quad \text{for }\quad \mathcal{E}_{d=1} \\
c_{\tau}(\lambda) &\simeq \frac{\sigma^{2}\ln(\lambda\tau)}{8\pi\lambda\tau} \quad \text{for }\quad \mathcal{E}_{d=2} \\
c_{\tau}(\lambda) &\simeq \frac{1.516 \, \sigma^{2}}{2\lambda\tau}  \quad \text{for }\quad \mathcal{E}_{d=3}\\
c_{\tau}(\lambda) &\simeq \frac{K-1}{(K-2)\lambda\tau} \quad \text{for }\quad \mathcal{T}_{K}
\end{align}

\section{Numerical simulations}

Simulating multiplicative stochatic equations like Eq.\ref{PAM} over a graph $\mathcal{S}$ can be challenging. We choose a split-step algorithm for the dynamic evolution that updates the solution in two stages. On site $i$ of the lattice, it reads:
\begin{align}
\nonumber
\phi \left(x_i, t+\frac{dt}{2} \right) &= \phi(x_i,t) \exp \left( \int_{t}^{t+dt}\nu_{\tau}(x_i,s)ds \right) \\ 
\label{Splitted_scheme}
\phi(x_i,t+dt)&= \phi(x_i,t+\frac{dt}{2}) + \frac{dt}{|\mathcal{N}(i)|}\left(\sum_{j\in \mathcal{N}(i)}\phi(x_j,t+\frac{dt}{2})-\phi(x_i,t+\frac{dt}{2})\right)
\end{align}
When $dt \rightarrow 0$, at first order in $dt$, we recover the usual discretization.
\begin{align}
\phi(x_i,t+dt)&= \phi(x_i,t)(1 + \nu_{\tau}(x_i,s) dt) + \frac{dt}{|\mathcal{N}(i)|} \sum_{j\in \mathcal{N}(i)} \phi(x_j,t)-\phi(x_i,t)
\end{align}

The advantage of using such split scheme is that one can compute exactly the statistics of $\int_{t}^{t+dt}\nu_{\tau}(x_i,s)ds$, whose description is given in \cite{PhysRevE.54.2084}, in order to remove one source of discretization error. The noise $\nu_{\tau}(x_i,s=0)$ is initially drawn from the stationnary distribution. 

At each time step, we renormalize the field $\lbrace \phi(.,t) \rbrace$ by its maximum value on the lattice $\hat{\phi}(t)=\max_{x_i \in \mathcal{S}}(\phi(x_i,t))$ and we store its logarithm in a variable, to keep track of these renormalization constants. This is necessary to keep the field from blowing up exponentially. We finally wait until convergence of the quantity $c_{\tau}(\lambda)= \lim_{t \rightarrow \infty}\frac{\langle \log(\phi(0,t))\rangle}{t}$ (usually for $t_{tot} \simeq 100$ and report the stationnary values. We repeat this procedure over several systems for the same lattice for different realisations of the noise and compute the averaged $c_{\tau}(\lambda)$ obtained, which is the one we report in the graphs of this paper. Note by constrat that $\bar{c}_{\tau}(\lambda)$ is very difficult to obtain numerically, as it does not average well.

The typical size of the lattices, both for Euclidean graphs or random regular graph, is of roughly $M=10^{6}$ sites. The typical number of noise realisations we used to average $c$ is $5$. The values of $\tau$ explored range from $0$ to $10$, of $\lambda$ from $10^{-3}$ to $10^2$.

\section{Methods for the financial dataset}

For the financial datasets, we employ two very different assets:
\begin{itemize}
\item The S\&P $500$ dataset we studied is provided free of charge by QuantQuote at the following web address: https://quantquote.com/historical-stock-data . It contains the daily value (or tick) of stocks dating back from 1998 and up to 2013. We focused on $M=388$ stocks who survived the entire period of measurements.
\item The government bonds yield dataset was also provided free of charge by OECD on the following website: https://data.oecd.org/interest/long-term-interest-rates.htm\#indicator-chart . It consists of measure on a monthly basis dating back from 1953 up to today. We retained $M=23$ countries, with a total number of $285$ months for our study. \\
\end{itemize}

Each dataset provides the tick value of those assets $p_i(t)$, from which we extract the log-return, analogous to our resources $r_i(t) = \log(p_i(t+1)/p_i(t))$. We then detrend the set, by removing the average of those returns, over the whole portfolio and period.
\begin{figure}[hbt!]
\includegraphics[width=1.0\textwidth]{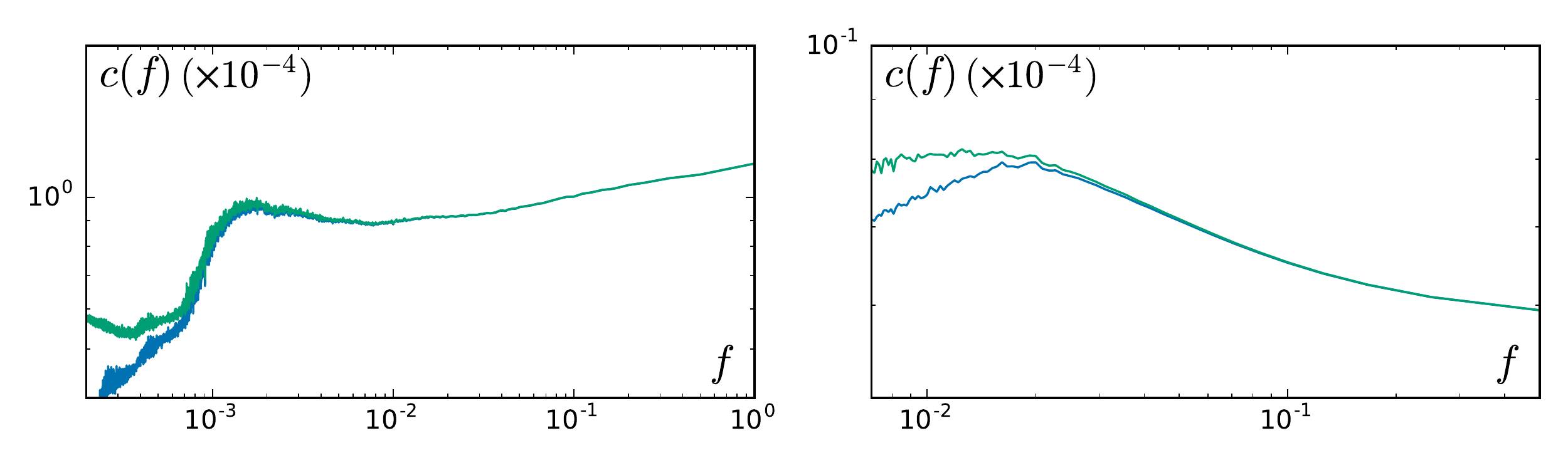} 
   \caption{Returns of the portfolios. $\mathbf{Left}$: S\&P dataset, the blue (lowest) curve is the return $c(f)$, while the green (upper) curve is the "annealed" return $\bar{c}(f)$.  $\mathbf{Right}$: Same curves for the government bonds dataset.}
   \label{empi_returns}
\end{figure}
\begin{figure}[hbt!]
\includegraphics[width=1.0\textwidth]{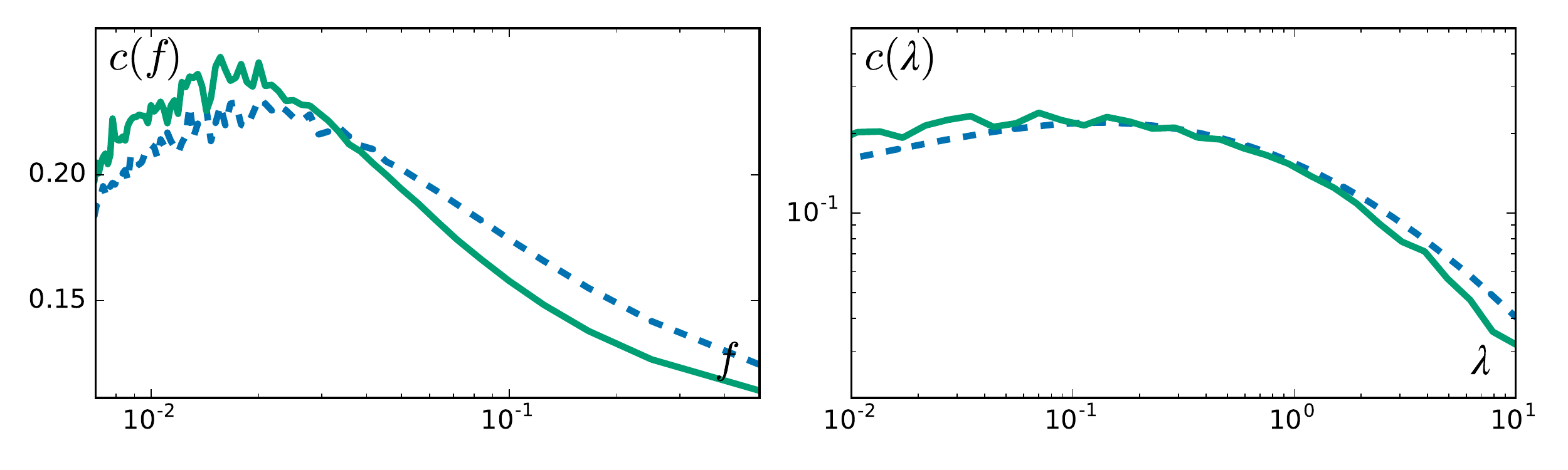} 
   \caption{Comparison of the return of the bonds portfolios and the growth rate of Eq.\ref{PAM}. $\mathbf{Left}$: portfolio return $c(f)$ for a rebalancing strategy over a fully connected graph (green, full) and over a random graph of degree $K=4$ (blue, dashed). $\mathbf{Right}$:  Growth rate $c_{\tau}(\lambda)$ again for a diffusion over a fully connected graph (green, full) and over a random graph of degree $K=4$ (blue, dashed)}
   \label{compa_empi}
\end{figure}

For both dataset, we apply the same simulation procedure. We first consider a graph, the nodes being the various assets (for the  S\&P, a fully connected graph, for the bonds both a fully connected graph and a random regular graph with $K=4$). We also start from an homogeneous state, with a small investment of $1$ on each node. We then simulated the evolution of the portfolio using the empirical returns $p_i(t)$.

With a frequency $f$, so after a period of $1/f$ months, we choose to rebalance our portfolio, according to its topology: the money on one site is splitted evenly among itself and its neighbours (for the complete graph, this rebalancing amounts to redistribute evenly the money on all countries). Thus, the rebalancing strength, although not diffusive, still scales with $f$. Finally, we measured both $c(f)=\frac{\langle \log(\phi_{i}(t))\rangle}{t}$ and $\bar{c}(f)=\frac{\log( \langle \phi_{i}(t) \rangle)}{t}$. The average is performed by choosing the starting point of the rebalancing strategy at random between $t=0$ and $t=1/f$, and summing over the last 20 steps.

We also compute the autocorrelation function $\text{ACF}(u)$ for a delay $u$ and averaged over the whole portfolio, defined as:
\begin{align}
\text{ACF}(u) = \frac{1}{M} \sum_{i \in \mathcal{S}} \sum_{t=1} ^{T} p_i(t)p_i(t+u)
\end{align}

The results of the rebalancing strategy over a fully connected portoflio are presented in Fig.\ref{empi_returns} for both datasets, and a similar plot can be found in the main text Fig.\ref{pos_hist}, along with the ACF in insets. 

Finally, we present, head to head, results from the analytical model Eq.\ref{PAM} and the empirical dataset in Fig.\ref{compa_empi}. Although the growth rate of the PAM $c_{\tau}(\lambda)$ and the return of the portfolio $c(f)$ are quite different, they strikingly behave in the same qualitative way, with respect to the structure of $\mathcal{S}$. Namely, the fully connected space allows the largest overall $c_{\tau}(\lambda_m)$ and $c(f_m)$, but decays faster in the delocalized phase compared to the same rates over $\mathcal{S}=\mathcal{T}_{K=4}$.
 
\end{document}